\documentclass[
superscriptaddress,
 reprint,
 amsmath,amssymb,
 aps, pra,
]{revtex4-2}   

\def\final{0}

\usepackage{verbatim}
\usepackage{rotating, booktabs}  
\usepackage{braket}
\usepackage{graphics}
\usepackage{graphicx}
\usepackage{hyperref}
\usepackage{color, xcolor}
\usepackage{upgreek}  
\usepackage{dcolumn}
\newcolumntype{z}[1]{D{.}{.}{#1}}
\usepackage{subfigure}
\usepackage{float}    

\usepackage{etoolbox}

\usepackage{comment}

 \usepackage{threeparttable}
 \usepackage{multirow}
 \usepackage{float}
 \usepackage{makecell}
 \usepackage{booktabs, tabularx, colortbl}
 \usepackage{array}
 \usepackage{enumerate}

\usepackage[linesnumbered,ruled ]{algorithm2e}
\SetKwInput{KwOutput}{Output}              
\SetKwInOut{Input}{Input}
\SetKwInput{Output}{Output} 

\usepackage{xargs}
\usepackage[colorinlistoftodos,prependcaption]{todonotes}

\newtheorem{definitionenv}{Definition}
\newtheorem{lemmaenv}[definitionenv]{Lemma}
\newtheorem{theoremenv}[definitionenv]{Theorem}
\newtheorem{corollaryenv}[definitionenv]{Corollary}
\newtheorem{propositionenv}[definitionenv]{Proposition}
\newtheorem{conjectureenv}[definitionenv]{Conjecture}
\newtheorem{remarkenv}[definitionenv]{Remark}
\newenvironment{remark}{\begin{remarkenv}\rm}{\end{remarkenv}}
\newcommand{\br}{\begin{remark}}
	\newcommand{\er}{\end{remark}}

\newtheorem{exampleenv}{Example}
\newtheorem{app-lemmaenv}[section]{Lemma}

\newenvironment{definition}{\begin{definitionenv}\rm}{\end{definitionenv}}
\newenvironment{lemma}{\begin{lemmaenv}\rm}{\end{lemmaenv}}
\newenvironment{theorem}{\begin{theoremenv}\rm}{\end{theoremenv}}
\newenvironment{corollary}{\begin{corollaryenv}\rm}{\end{corollaryenv}}
\newenvironment{example}{\begin{exampleenv}\rm}{\end{exampleenv}}
\newenvironment{proposition}{\begin{propositionenv}\rm}{\end{propositionenv}}
\newenvironment{conjecture}{\begin{conjectureenv}\rm}{\end{conjectureenv}}
\newenvironment{app-lemma}{\begin{app-lemmaenv}\rm}{\end{app-lemmaenv}}

\newcommand{\bd}{\begin{definition}}
	\newcommand{\ed}{\end{definition}}
\newcommand{\bl}{\begin{lemma}}
	\newcommand{\el}{\end{lemma}}
\newcommand{\elp}{\hspace*{\fill} $\Box$
\end{lemma}}
\newcommand{\bt}{\begin{theorem}}
\newcommand{\et}{\end{theorem}}
\newcommand{\etp}{\hspace*{\fill} $\Box$
\end{theorem}}
\newcommand{\bc}{\begin{corollary}}
\newcommand{\ec}{\end{corollary}}
\newcommand{\ecp}{\hspace*{\fill} $\Box$
\end{corollary}}
\newcommand{\bcj}{\begin{conjecture}}
\newcommand{\ecj}{\end{conjecture}}

\newcommand{\be}{\begin{example}}
\newcommand{\ee}{\end{example}}
\newcommand{\eep}{\hspace*{\fill} $\Box$
\end{example}}
\newcommand{\bp}{\begin{proposition}}
\newcommand{\ep}{\end{proposition}}
\newcommand{\epp}{
\end{proposition}}

\newcommand{\cS}{\mathcal{S}}

\newcommand{\TB}{\text{LOOKUP}}

\ifnum\final=0
\newcommand{\mynote}[2]{{\color{#1} \marginpar{\tiny #2}}}
\newcommand{\mybignote}[2]{{\color{#1} $\langle \langle$ #2$\rangle \rangle$}}
\newcommandx{\rednote}[2][1=]{\todo[linecolor=red,backgroundcolor=red!25,bordercolor=red,#1]{#2}}
\newcommandx{\bluenote}[2][1=]{\todo[linecolor=blue,backgroundcolor=blue!25,bordercolor=blue,#1]{#2}}
\newcommandx{\yellownote}[2][1=]{\todo[linecolor=yellow,backgroundcolor=yellow!25,bordercolor=yellow,#1]{#2}}
\newcommandx{\greennote}[2][1=]{\todo[inline,linecolor=olive,backgroundcolor=green!25,bordercolor=olive,#1]{#2}}

\else
\newcommand{\mynote}[2]{}
\newcommand{\mybignote}[2]{}
\newcommand{\rednote}[2][1=]{}
\newcommand{\bluenote}[2][1=]{}
\newcommand{\greennote}[2][1=]{}
\newcommand{\yellownote}[2][1=]{}

\fi

\begin{document}

\title{  Parallel   syndrome extraction with shared flag qubits for Calderbank-Shor-Steane codes of distance three}

\author{Pei-Hao Liou}
 \email{phliou.ee08@nycu.edu.tw}
 \affiliation{Institute of Communications Engineering, National Yang Ming Chiao Tung University,  Hsinchu 30010, Taiwan} 
\author{Ching-Yi Lai}%
 \email{cylai@nycu.edu.tw}
 \affiliation{Institute of Communications Engineering, National Yang Ming Chiao Tung University,  Hsinchu 30010, Taiwan}
 \affiliation{Physics Division, National Center for Theoretical Sciences, Taipei 10617, Taiwan}

\date{\today}

 \begin{abstract}
 To perform achieve fault-tolerant quantum computation, one can use flagged syndrome extraction  with fewer ancilla qubits.
 However, it suffers from long circuit depth if one stabilizer is measured at a time.
 Previously, Reichardt showed that it is possible to measure multiple stabilizers  with at most one shared flag qubit for certain small quantum codes. In this paper, we propose a procedure for general Calderbank-Shor-Steane codes of distance three so that multiple  $Z$-stabilizers
 ($X$-stabilizers) can be fault-tolerantly measured in parallel with one shared flag qubit.  
 We simulate the memory and computation pseudo-thresholds for various code schemes. In particular, our parallel scheme based on Shor's nine-qubit code performs better than known seven- and nine-qubit schemes in the literature.

 
 \end{abstract}

\maketitle

\section{Introduction}
Physical platforms for quantum computing usually have finite qubit coherent time and high gate error rates \cite{takita17,Vui18,postler2021,Egan2021}.
To have reliable quantum computers,  it is necessary to  realize fault-tolerant quantum computation (FTQC)~\cite{AB97,Shor96,DS96,KLZ96,Got97,Pre98c,Ste97L,Ste99N}.  
The central idea of FTQC is to compute directly on quantum states encoded in a quantum error-correcting code. Encoded logical operations are implemented and quantum error corrections are constantly applied, so that an arbitrarily accurate quantum computation is possible when the error rates of physical gates are below a certain threshold~\cite{AB97}. 
 
To perform a fault-tolerant syndrome extraction for error correction, one may use one of the Shor, Steane, and Knill syndrome extraction procedures~\cite{Shor96,Ste97L,Knill05}.
Usually repeated syndrome measurements are performed with the Shor syndrome extraction~\cite{Shor96} to obtain a reliable error syndrome. One may also design quantum data-syndrome codes to handle syndrome measurement errors~\cite{ALB20,KCL21}. 
If the Steane or Knill syndrome extraction is used, no repeated measurements are required since one  has to handle only an effective error and a residual error will be treated in the next error correction cycle~\cite{ZLB+20}. 
However, large ancilla blocks are required for  the Steane or Knill syndrome extraction.
Huang and Brown recently proposed another syndrome extraction method, which is a compromise of both the Shor and the Steane extractions~\cite{HB21}. 

It is also possible to perform error correction for certain quantum codes with direct syndrome extractions~\cite{FMMC12,LMB18}
but usually a decoding strategy is required, such as repeated syndrome measurements followed by a perfect matching.

 In \cite{Yoder2017}, Yoder and Kim showed that fewer ancilla qubits are required for FTQC based on certain quantum codes. Following that, Chao and Reichardt introduced the idea of flag qubits for syndrome extraction so that fewer ancilla qubits are required for a syndrome extraction procedure to be fault-tolerant~\cite{CR18a,CR18b}.
 They also showed  that multiple stabilizers can  be measured with one shared flag qubit for certain small codes.
The central idea is to enlarge the space of syndrome bits by introducing additional flag qubits and controlled-Not (CNOT) gates such that any single location failure will have a unique error syndrome  and thus can be corrected. Since a flag FTQC scheme requires fewer qubits, it is suitable for near-term quantum devices. 
More general flag schemes have been studied for FTQC based on stabilizer codes and magic state distillation \cite{CB18,fault_magic,TCL20,CR20}.

A standard flagged syndrome extraction circuit measures a single stabilizer generator at a time.
If a complete syndrome extraction circuit is composed of several flagged syndrome extraction circuits in a cascade form, most qubits are idle  most of the time. When the error rate for idling is high, the benefit of using fewer ancilla qubits with the flagged syndrome extraction may be compensated.  To handle this problem of long circuit depth, Reichardt proposed the idea of  flagged syndrome extraction for multiple stabilizers with shared flag qubits for certain small codes of distance three~\cite{Rei20}. Moreover, he showed that two or three stabilizer of the [[7,1,3]] code~\cite{Ste96} can be fault-tolerant measured in parallel without additional flag qubits.
(Reichardt's protocol is reviewed in Appendix~\ref{sec:parallel_flag}.)

In this paper,  we propose a procedure (Algorithm~\ref{algorithm:shared_flag}) for parallel syndrome extraction with shared flag qubits for a general Calderbank-Shor-Steane (CSS) code of distance three.
	We show that multiple $Z$-stabilizers can be measured with one shared flag qubit and similarly for multiple $X$-stabilizers.
	In this way, the circuit depth can be greatly reduced compared to the standard flagged syndrome extraction.
	In particular, we show that all the $Z$-stabilizers ($X$-stabilizers) can be measured in parallel with one flag qubit in the scenario of fault-tolerant error detection.	

In addition, we show that for flagged syndrome extraction of a CSS code, a complete unflagged syndrome extraction is not necessary when a flag rises.
More precisely, we observe that in a parallel syndrome extraction of multiple $Z$-stabilizers,
the residual errors when a flag rises will have at most a weight-one Pauli $X$ error, which can be corrected in the next error correction cycle,
so only $X$-stabilizers need to be measured to catch high-weight $Z$ errors. In this way, the circuit depth can be further reduced. 
Accordingly, we propose Algorithm~\ref{alg:dec_condition_CSS} that decodes flagged scheme of CSS codes in an adaptive way.

	We remark that   one $X$-stabilizer and one $Z$-stabilizer (two dual operators) can be measured in parallel with a shared flag qubit using Reichardt's method for the [[15,7,3]] code. On the contrary, our procedure applies to only stabilizers of the same type. Thus our procedure applies to asymmetric  CSS codes with unequal numbers of $X$- and $Z$- stabilizers, such as the [[15,1,3]] Reed-Muller code.

Finally, we simulate and compare several flag and parallel schemes based on the [[4,2,2]], [[7,1,3]], and Shor's [[9,1,3]]~\cite{Shor95} codes
for error detection and error correction, respectively.
Both the memory and computation pseudo-thresholds are simulated for each scheme. 
Note that we do not assume  a two-dimensional layout for a FTQC scheme with physical restrictions~\cite{SDT07,SR09,LPSB13}
and nonlocal CNOTs are allowed.

We  observe that a parallel scheme will perform better than its nonparallel counterpart.
In particular, our [[9,1,3]] parallel scheme has the memory pseudo-thresholds of
$9.82 \times 10^{-3}$  when the memory is error-free  
and $8.84 \times 10^{-4}$ when all the gate and memory error rates are the same.
These numbers are better than those obtained by  Reichardt's [[7,1,3]] parallel scheme (see Table~\ref{tb:std_result}).

We also compare our [[9,1,3]] parallel scheme with known nine-qubit code schemes in the literature. 
In~\cite{LMB18}, Li et al. proposed a fault-tolerant quantum memory scheme for the [[9,1,3]] Bacon-Shor code using only four ancilla qubits,
called  \textit{Bacon-Shor-13} and they showed this scheme outperforms a scheme based on the [[9,1,3]] surface codes using eight ancilla qubits. From the circuit design, one can observe that Bacon-Shor-13 performs better when the memory rate is low since it has fewer qubits involved (a total of 13 qubits).  

As can be seen in Table~\ref{tb:std_result}, our [[9,1,3]] parallel scheme outperforms Bacon-Shor-13 in  both the memory and the computation pseudo-thresholds.

This paper is organized as follows. 
In Section \ref{sec:intr_flag}, we briefly introduce the basics of stabilizer codes, FTQC, and flagged syndrome extraction.  
In Section \ref{sec:flag_sharing}, we show how to perform parallel syndrome extraction for multiple stabilizers of a CSS code with one shared flag qubit.
Simulations our parallel schemes are provided and compared in Section \ref{sec:sim}.
Then  we conclude in Section \ref{sec:conclusion}.

\section{Preliminaries}\label{sec:intr_flag}
  
 \subsection{Stabilizer Codes}
  We consider quantum errors that are tensor product of Pauli matrices  $I=\left(\begin{array}{cc}
  	1 & 0\\	0 & 1  
  \end{array}\right)$, $X=\left(\begin{array}{cc}
  0 & 1\\
  1 & 0
\end{array}\right)$, $Y=\left(\begin{array}{cc}
0 & -i\\
i & 0
\end{array}\right)$, and $Z=\left(\begin{array}{cc}
  	1 & 0\\
  	0 & {-1}
  \end{array}\right)$ in the computational basis \{$|0\rangle$, $|1\rangle$\}.
For convenience, we omit the symbol of tensor product for $n$-fold Pauli operators. 
In addition, we may sometimes omit the identity $I$ and use an index to indicate which qubit a Pauli matrix operates on. For example, $X\otimes Y\otimes Z\otimes I\otimes I$ will be denoted as  $X_1Y_2Z_3$.
An $X$- or $Z$-type Pauli operator  has nonidentity components that are all $X$ or all $Z$ matrices. 
For an $X$- or $Z$-type Pauli operator,  the redundant $X$s or $Z$s will be suppressed. 
For example, $X_1X_2X_3$ may also be denoted as $X_{1,2,3}$.
The \textit{weight} of an $n$-fold Pauli operator is the number of its nonidentity elements. 
 Usually a high-weight error is less likely than a low-weight error in an error model.
Consequently, we would like to handle the more-likely events in error correction.

An $[[n,k,d]]$ stabilizer code $C(\cS)$  is defined as the joint $+1$ eigenspace of an Abelian subgroup  $\cS$ of the $n$-fold Pauli operators with $n-k$ independent generators $g_1,\dots,g_{n-k}$ 
and the operators in $\cS$ are called \textit{stabilizers}.
It encodes $k$ logical qubits into $n$ physical qubits and can detect errors up to weight $d-1$
or correct errors up to weight $\lfloor\frac{d-1}{2}\rfloor$~\cite{Got97}.

\bd
The \textit{error syndrome} of a Pauli error $E$  with respect to the $n-k$ stabilizer generators $g_i\in\cS$ is a binary string of length $n-k$, whose $i$-th bit  is $0$ if $E$ commutes with $g_i$, and $1$, otherwise.
\ed
If an error has a nonzero error syndrome, it can be detected and possibly corrected.
\bp
If each operator of a set of Pauli errors  has a unique error syndrome with respect to a stabilizer group, then  
this set of Pauli errors is correctable.
\ep
Clearly one can build a \textit{lookup table} that consists of each  error operator and its error syndrome for error correction~\cite{MS77}.

A CSS code is a stabilizer code defined by $X$-type  or $Z$-type stabilizer generators~\cite{CS96,Ste96a}.
$X$ and $Z$ errors can be separately treated in a CSS code.
A CSS code of distance three can correct a single $X$ error and a single $Z$ error, simultaneously.

 \subsection{Fault-tolerant quantum computation}

 In a quantum circuit, quantum errors occur due to imperfect quantum gates or memory and they may propagate through multiple-qubit gates.
 \bd
 A \textit{location failure} is referred to as an event that a Pauli error occurs after a perfect single-qubit gate, two-qubit gate, ancilla preparation, qubit measurement, or an idle qubit in a circuit. 
 \ed
  To perform quantum computation with imperfect gates and memory, 
 quantum information has to be encoded in a quantum stabilizer code, and encoded quantum operations are performed 
 so that quantum error correction is constantly performed.

 \bd
 A  location in a procedure is called a \textit{bad location} if one of  its location failures  may evolve into an 
 uncorrectable (respectively, undetectable) error  for a code of distance three (respectively, two).
 Such location failures are called \textit{bad location failures}.
 \label{def:Bad_loc}
 \ed
 
\bd
A procedure is called \textit{fault-tolerant} if there are no bad locations in the procedure. 
\ed

\begin{figure}[htbp]
 	\centering
 	\includegraphics[width=0.9\columnwidth]{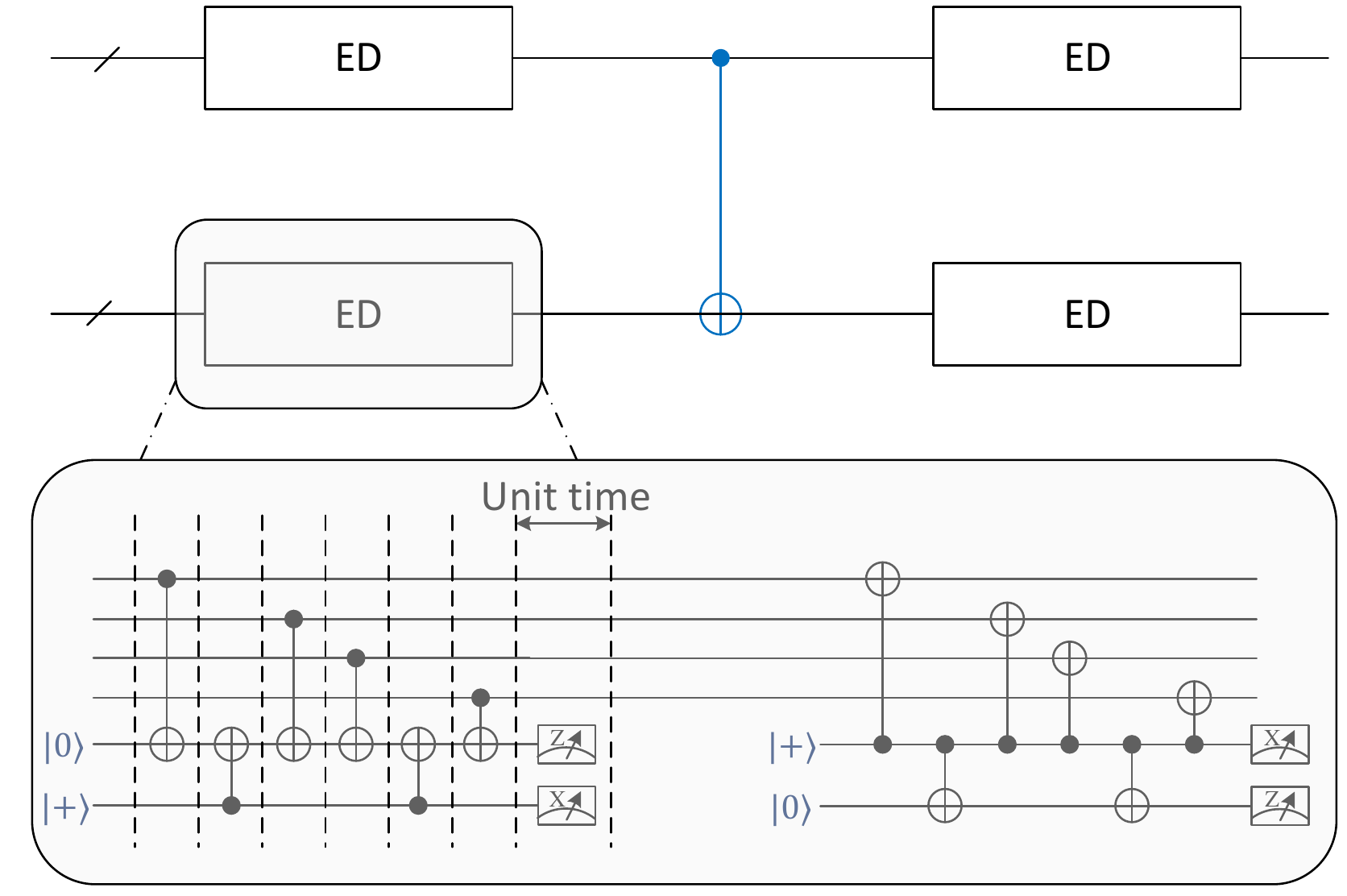}
 	\caption{The ex-Rec CNOT  for the {[[4,2,2]]} code with flagged syndrome extraction. 
 			Two quantum codewords encoded by the $[[4,2,2]]$ code are first processed by two error detection (ED) blocks.
 			Then bitwise CNOT gates (colored in blue) are applied, followed by another two ED blocks.   			
 			A fault-tolerant ED block circuit measures the two stabilizers $ZZZZ$ and $XXXX$ for the $[[4,2,2]]$ code.
 			In this example, flagged syndrome extraction is used, which is explained in Fig.~\ref{fig:fault_tolerant measurment_Z}.   
 		    }\label{fig:Logic CNOT_flag}
 \end{figure}

 For example, a fault-tolerant circuit of a CNOT gate on two qubits is encoded to the circuit in Fig.~\ref{fig:Logic CNOT_flag}
 where each qubit is encoded into a CSS codeword and bitwise CNOT gates are applied. Note that an error detection (ED) or error correction block is performed to each codeword prior to  and subsequent to the bitwise CNOT gates. 
 Such implementation of a logical CNOT gate is  called \textit{an extended rectangle (ex-Rec)} of the CNOT gate~\cite{AGP06}.

\bd
 The \textit{error threshold} for a procedure is the (physical) error rate below which the logical error rate of the procedure would be lower than the error rate.
 \ed
 
 We consider two types of thresholds: \textit{memory threshold} and \textit{computation threshold}.
 The memory threshold is the error threshold for error correction or detection on a quantum codeword.
 The computation threshold  is the error threshold for an ex-Rec CNOT gate since the ex-Rec CNOT is usually the most complicated procedure in a universal set of gates.

 Simulations in this paper assume the following error model.
 We add independent depolarizing errors with  rate $p$ as quantum operations
 after gates or before measurements in the circuit. 
 Assume that each gate takes the same unit of time. 
 Each single-qubit location undergoes $X$, $Y$, or $Z$ with probability $p/3$
 and each qubit-measurement is corrupted with probability $2p/3$. Each CNOT gate is followed by one
 of the 15 non-identity two-qubit Pauli operators with probability $p/15$.
 An idle qubit suffers depolarizing errors with rate $\gamma p$ for  $0\leq \gamma\leq 1$.

\subsection{Fault-tolerant syndrome extraction with flag qubits}

In general a syndrome extraction circuit has many bad locations. By introducing additional ancillary qubits, called \textit{flag qubits}, one can carefully place  controlled-NOT (CNOT) gates so that
bad location failures will trigger these flag qubits, and thus, be detected~\cite{CR18a,CR18b}.

\begin{figure}[htbp]
	
		\centering
	
			\includegraphics[width=1\linewidth]{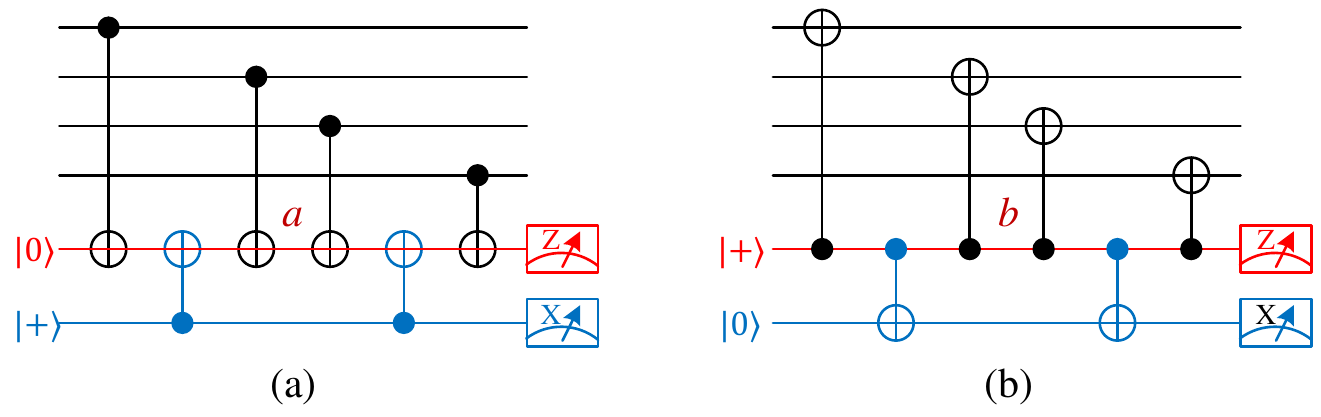}

		\caption{(a) Flagged fault-tolerant syndrome extraction for $Z_1Z_2Z_3Z_4$.
			The ancilla qubit initialized in $\ket{+}$ is a flag qubit, which  is coupled to the raw syndrome measurement circuit via two additional CNOTs.
			(b) Flagged fault-tolerant syndrome extraction for $X_1X_2X_3X_4$. }\label{fig:fault_tolerant measurment_Z}
	
\end{figure}

Figure~\ref{fig:fault_tolerant measurment_Z}~(a) illustrates how  a weight-4 stabilizer  $Z_1Z_2Z_3Z_4$ is measured.
Four CNOT gates  are used to check the parity of four data qubits and the result is left in the first ancilla qubit (marked in red), which is initialized in $\ket{0}$.
Observe that location $a$ in Fig.~\ref{fig:fault_tolerant measurment_Z}~(a)  is a bad location since a $Z$ error here  will induce a weight-2 $Z$ error on the data qubits. 
This event cannot be detected by the stabilizer measurements but it will be detected using a flag qubit with elegantly positioned CNOT gates (marked in blue) as shown in~Fig.~\ref{fig:fault_tolerant measurment_Z}~(a).
We say that this failure triggers the flag qubit and the flag rises.
This method is called \textit{flagged syndrome extraction}~\cite{CR18a,CR18b}.
The syndrome extraction circuit without the flag qubit and the two CNOT gates will be called an \textit{unflagged syndrome extraction circuit}.

\begin{figure}[htbp]
	\begin{minipage}[t]{0.48\textwidth}
		\centering
	 \includegraphics[width=0.9\linewidth]{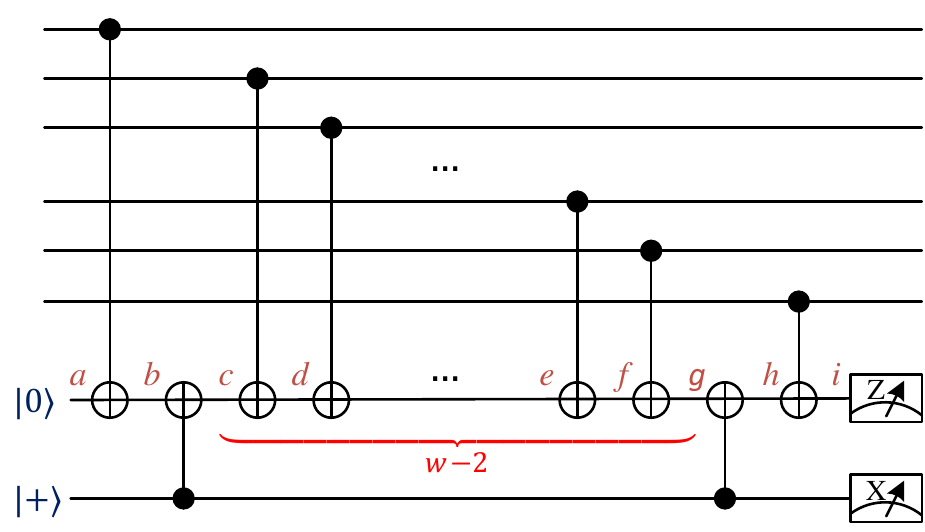}
		\caption{ Standard flagged syndrome extraction for a $Z$-type stabilizer of weight $w$.
		  }\label{fig:flagged_syndrome_extraction}
	\end{minipage}
\end{figure}

 The measurement of a weight-4 stabilizer $X_1X_2X_3X_4$ is similarly implemented in Fig.~\ref{fig:fault_tolerant measurment_Z}~(b). 
 These circuits can be directly generalized for measuring $Z$- or $X$-stabilizers of higher weight as shown in Fig.~\ref{fig:flagged_syndrome_extraction}.
This is called a \textit{standard flagged syndrome extraction circuit}. 
 A circuit for measuring the error syndrome with respect to a set of stabilizer generators will be called a \textit{complete} syndrome extraction circuit.

In the following, we discuss how flagged syndrome extraction is used in fault-tolerant quantum error detection and correction.

\subsubsection{Quantum error detection with flag qubits}
 
A fault-tolerant error-detection scheme has to be designed such that a detectable error will not evolve into an undetectable error. 
A flagged syndrome extraction circuit ensures that any bad location failure will trigger a flag qubit and be detected.
Therefore, one learns that some error occurs if anyone of the measurement outcomes in the flagged syndrome extraction circuits is $-1$, and the codeword has to be discarded.

   Consider the  $[[4,2,2]]$ error-detecting code \cite{Got97}, which is defined by two stabilizer generators 
$X_1X_2X_3X_4$ and $Z_1Z_2Z_3Z_4.$ Its complete syndrome extraction consists a sequential implementation of Figs.~\ref{fig:fault_tolerant measurment_Z}~(a) and~\ref{fig:fault_tolerant measurment_Z}~(b)
as shown in Fig.~\ref{fig:Logic CNOT_flag}.
This procedure is fault-tolerant.

\subsubsection{Quantum error correction with flag qubits}
In a fault-tolerant error correction scheme with flagged syndrome extraction, error recovery is adaptively performed according to the measured error syndrome and flag qubits. The basic error correction rules for distance-three quantum codes are as follows. 

A complete flagged syndrome extraction circuit is first implemented.

If no flag qubits rise, it is most likely that no bad location failure occurs, and a usual syndrome decoding is performed according to the measured error syndrome. If one flag qubit rises during the complete syndrome extraction, a round of complete unflagged syndrome extractions will be performed right away, and error recovery is performed according to  the flag qubit as well as its measured error syndrome.

When a flag qubit rises, we know that there is some high-weight error in the codeword. In this case, the conventional minimum-weight decoding rule does not work, and we are not aware of any general decoding algorithm. 
We have to choose the most likely error according to the flag qubits and the error syndrome from a complete unflagged syndrome extraction.

	Let $f\in\{0,1\}^r$ be the measurement outcomes on the flag qubits in a complete flagged syndrome extraction circuit with $r$ flag qubits.
We denote  \TB$(f)$ as the syndrome look-up table according to the flag   outcomes $f$.
	Note that $\TB(0)$ is the original syndrome table.

The standard decoding procedure for flagged syndrome extraction in Ref.~\cite{CR18a} is summarized in  Algorithm~\ref{algorithm:standard}. 

\begin{algorithm}[htp]\label{algorithm:standard}
	
	Suppose that there are $r$ stabilizer generators $g_1,g_2,\ldots,g_r$ and $f$ is initialized to $0^r$\; 
	
	\For{$i= 1,2,\ldots,r$}{
		Apply $C(g_i)$ and get syndrome  bit $m_i$\;
		Update $f_i$ according to the measurement outcome on the flag qubit in $C(g_i)$\;
		
		\uIf{$f_i=1$ }{
			Apply a complete unflagged syndrome extraction circuit to get syndrome bits $m'_{ }\in\{0,1\}^r$\;
			Perform error correction according to $\TB(f)$ and $m'$\;
			break\;
	 }

		\ElseIf {$m_i=1$ }{ 
		   	Apply a complete unflagged syndrome extraction circuit to get syndrome bits $m'_{}\in\{0,1\}^r$\;
			Perform error correction according to $\TB(0)$ and $m'$\;
			break\;					
			}
	}

	\caption{ Standard decoding procedure for flagged syndrome extraction. }
\end{algorithm}

\section{flag-shared syndrome extraction   for codes of distance two or three }\label{sec:flag_sharing}

In this section, we propose a procedure for parallel syndrome extraction with shared flag qubits for a general CSS code of distance three.
We begin with the notation. 

\subsection{Parallel flagged syndrome extraction with one shared flag qubit}
Suppose that there are $r$ stabilizer generators $g_1,\dots,g_r$ to be measured for an  [[$n,k,3$]] CSS code.
Let  $C(g_i)$ be a (standard) flagged syndrome extraction circuit for $g_i$ as shown in Fig.~\ref{fig:flagged_syndrome_extraction} for a $Z$-stabilizer.
The flagged syndrome extraction circuit is designed so that each bad location failure will trigger the flag qubit. 
Let $B(C(g_i))$ denote the number of location failures in $C(g_i)$ that will trigger the flag qubit.

 An observation is that $B(C(g_i))$ depends on the weight of the stabilizer $g_i$.
\bp \label{pro:count}
Suppose that $g_i$ is an $X$- or $Z$-stabilizer of weight $w$.
Then $B(C(g_i))\leq w-1.$
\ep
\noindent\textbf{Proof. }
Consider a $Z$-stabilizer $g_i$ of weight $w$ with a standard flagged syndrome extraction circuit $C(g_i)$ as shown in Fig.~\ref{fig:flagged_syndrome_extraction}. 
There are two CNOTs connecting the flag and the ancillary qubits. Other $w-2$ CNOTs are placed in sequence between these two CNOT gates.
All the location failures at $a$, $b$, $h$, and $i$	will not trigger the flag qubit, whereas a Pauli $Z$ (or $Y$) at the other $w-1$ locations $c$, $d$, up to $g$ will. 

Note that a $Z$ error at $g$ is not a bad failure since it remains to be a weight-one error at the end of the circuit. 
The other $w-2$ location failures may evolve to high-weight $Z$ errors, which are potential bad failures.\\

\begin{figure*}[htbp]
	\centering
	\includegraphics[width=0.98\linewidth]{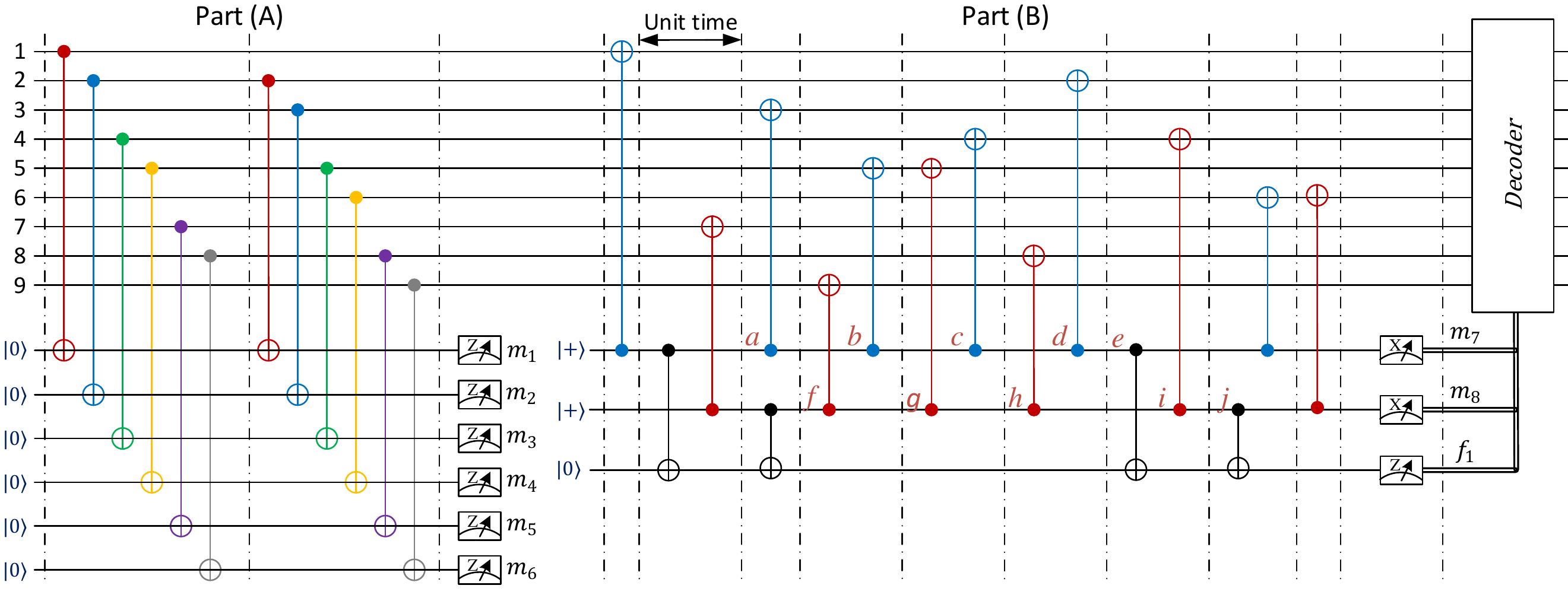}
	\caption{Parallel flagged syndrome extraction circuit for Shor's [[9,1,3]] code with one shared flag qubit.
		Part(A) extracts the syndrome bits for $Z_1Z_2$ (red), $Z_2Z_3$ (blue), $Z_4Z_5$ (green), $Z_5Z_6$ (yellow), $Z_7Z_8$ (purple) and $Z_8Z_9$ (gray). Part (B) extracts the syndrome bits for  $X_1X_2X_3X_4X_5X_6$ (blue) and $X_4X_5X_6X_7X_8X_9$ (red).}\label{fig:913_parallel}
\end{figure*}

 Another observation is that flagged syndrome extraction circuits for stabilizers of the same type
 may be joined together and  a shared flag qubit may be sufficient for fault-tolerant syndrome extraction. 
 For example, Fig.~\ref{fig:913_parallel}\, part (B) shows the joint flagged syndrome extraction circuit with one shared flag qubit for two stabilizers.

 Moreover, we find that no additional bad location failures are introduced in this joint circuit. Consider the two CNOTs connecting an ancilla qubit and the flag qubit. 
 A $Z$ error on the flag qubit prior to the first CNOT will be absorbed to the flag qubit,
 An $X$ error on the flag qubit will not propagate to any other ancillas, and it triggers the flag qubit.
 A $Z$ error on the flag qubit following the second CNOT has no effect on the flag qubit.
 An $X$ error on the flag qubit will not propagate to any other ancillas, and it triggers the flag qubit.
 No errors on one ancilla could propagate through the flag qubit to another ancilla because of the directions of CNOTs. Hence, we have the following proposition.
\bp  \label{pro:invariant}
Suppose that $g_{j_1}, \dots, g_{j_s}$ are stabilizers of the same type ($X$ or $Z$).
Suppose that a flag qubit is shared by $C(g_{j_1}), \dots, C(g_{j_s})$ and the CNOTs connecting the ancillas and the flag qubit are placed in sequence. Then a bad location failure in this joint circuit must have appeared in one of the $C(g_{j_i})$s; no additional bad location failure is introduced in the joint circuit.  
\ep

The next question is: how many stabilizer generators can be fault-tolerantly measured in parallel with only one shared flag qubit? 
Suppose that there are $x$ $X$-type stabilizers and $z$ $Z$-type stabilizers.
If $g_i$ is a $Z$-stabilizer, we observe that in  $C(g_i)$, a bad location failure  always evolves to a high-weight $Z$ error. 
For example, in Fig.~\ref{fig:fault_tolerant measurment_Z}\,(a) a $Z$ error at location $a$ evolves to $Z_3Z_4$ at the end of the circuit.
When a flag rises, this high-weight $Z$ error will be identified by measuring the $X$-stabilizers in an unflagged syndrome extraction circuit. Consequently, we must have $B(C(g_i))\leq 2^x$ so that all the location failures that trigger the flag qubit can be identified using an unflagged syndrome extraction. Similar for the flagged syndrome extraction of an $X$-stabilizer.

Consequently, we must have the following proposition for a flag-shared syndrome extraction scheme of multiple stabilizers of the same type, following Proposition~\ref{pro:invariant}.

 \bp
 Suppose that $g_{j_1}, g_{j_2},\dots,$ $g_{j_s}$ are stabilizers  of the same type and $C(g_{j_1}),$ $C(g_{j_2}),\ldots,$ $C(g_{j_s})$ can be performed in parallel using a shared flag qubit. 
Then the following conditions hold. 
 \begin{enumerate}
 
 		\item Each distinct bad location failure has a unique error syndrome when a flag rises.
 		\item 
 		
$\displaystyle
\begin{cases}
\sum_{i=1}^s B(C(g_{j_i})) \leq 2^z,& \mbox{ if $g_{j_1},\dots, g_{j_s}$ are $X$-type}; \\
\sum_{i=1}^s B(C(g_{j_i})) \leq 2^x,& \mbox{ if $g_{j_1},\dots, g_{j_s}$ are $Z$-type}, \\
	\end{cases}
$ 		
assuming that there are $x$ $X$-type stabilizers and $z$ $Z$-type stabilizers. 
 
 \end{enumerate}
 \label{pro:condition_parallel}
 
 \ep

According to Propositions~\ref{pro:count}, \ref{pro:invariant}, and \ref{pro:condition_parallel}, we have a procedure (Algorithm~\ref{algorithm:shared_flag}) for finding a parallel flagged syndrome extraction circuit for a set of stabilizers with a shared flag qubit.

\begin{algorithm}[htp]\label{algorithm:shared_flag}
 
	 Choose stabilizers $g_{j_1},g_{j_2},\ldots, g_{j_s}$ of $Z$ ($X$) type  such that $\sum_{i=1}^s B(C(g_{j_i})) \leq 2^x$ $(2^z)$,
		where $x$ $(z)$ is the number of independent $X$ $(Z)$ stabilizer generators\;
		
	  Let $C(g_{j_1}),C(g_{j_2}),\ldots, C(g_{j_s})$ share one flag qubit\; Let ${unique}=false$\;
		
		\While{\text{unique}}{
			randomly adjust the orders of a subset of CNOTs in 	 $C(g_{j_1}),C(g_{j_2}),\ldots, C(g_{j_s})$\;	
			
		\If{each of the locations that trigger the flag qubit has a unique error syndrome from an unflagged syndrome extraction\;}
		{let ${unique}=true$\;}
	}
	\caption{ Finding a parallel flagged syndrome extraction circuit with one shared flag qubit.}
\end{algorithm}

We remark that the number of possible error syndromes is exponential in the number of stabilizer generators, whereas the number of locations that trigger the flag qubit is linear in the number of stabilizer generators in parallel. As a consequence, this randomized algorithm is good enough when $x$ or $z$ is sufficiently large. 
We provide examples of Shor's [[9,1,3]] code and the [[15,1,3]] (punctured) Reed-Muller code in the following subsections.

We can extend the above idea for error detection.  Note that the conditions for error detection is that any bad location failure can be detected.  Since a bad location failure will trigger the flag qubit in a flagged syndrome extraction circuit, we have the following result.
\bl
All the $Z$-stabilizers of a  CSS code of distance two or three can be fault-tolerantly measured in parallel with a shared flag qubit for error detection.
Similarly for the $X$-stabilizers.  
\el

\subsubsection{Shor's [[9,1,3]] code	}\label{sec:913}
Shor's [[9,1,3]] code is constructed by concatenating a three-qubit phase-flip code with  a  three-qubit bit-flip QECC and can correct an
arbitrary $X$ error and an arbitrary $Z$ error simultaneously \cite{Shor95}. It is defined by stabilizers
\begin{equation}\label{eq:913_stabilizers}
\begin{aligned}
g_1=&Z_1Z_2,\,g_2=Z_2Z_3,\,g_3=Z_4Z_5,&\\
g_4=&Z_5Z_6,\,g_5=Z_7Z_8,\,g_6=Z_8Z_9,&\\
g_7=&X_1X_2X_3X_4X_5X_6,&\\
g_8=&X_4X_5X_6X_7X_8X_9.&
\end{aligned}
\end{equation}
Since each of $g_1,g_2,g_3,g_4,g_5$ and $g_6$  has weight two, its syndrome extraction circuit consists of two CNOTs.
 One can check that it has no bad location and no flag qubit is required. They can be measured in parallel as shown in Part (A) of Fig.~(\ref{fig:913_parallel}).
  
 The two stabilizers $g_7$ and $g_8$ can be fault-tolerantly measured using a parallel flagged syndrome extraction circuit by Algorithm~\ref{algorithm:shared_flag}.
 There are five location failures in $C(g_7)$ (or $C(g_8)$) that trigger the flag qubits, which satisfies Proposition~\ref{pro:count} with equality. 
 Since there are six $Z$-stabilizers, Proposition~\ref{pro:condition_parallel} holds with  $10< 2^6=64$.
 Consequently, a parallel syndrome extraction circuit can be easily found  as shown in Part (B) of Fig.~(\ref{fig:913_parallel}).
 
By Lemma~\ref{lemma:CSS_error} and Algorithm~\ref{alg:dec_condition_CSS} in the next subsection, the unflagged syndrome extraction circuit consists of solely Part (A) of Fig.~\ref{fig:913_parallel}. 
 The location failures are  listed in Table~(\ref{table:error_list_913_p}), together with their error syndromes 
 from a following separate unflagged syndrome extraction circuit.

   \begin{table}[htbp]
 	\begin{tabular}{|c|c|c|c|}
 		\hline
 		Failure  & Data error & $m_7m_8f_1$  & $m^{\prime}_{123456}$    \\ 
 		\hline
 		$X_a$ & $X_{2,3,4,5,6}$ & 001 & 100000  \\
 		$X_b$ & $X_{2,4,5,6}$   & 001 & 110000  \\
 		$X_c$ & $X_{2,4,6}$     & 001 & 111100  \\ 
 		$X_d$ & $X_{2,6}$       & 001 & 110100  \\  
 		$X_e$ & $X_6$           & 001 & 000100  \\
 		$X_f$ & $X_{4,5,6,8,9}$ & 001 & 000010  \\  
 		$X_g$ & $X_{4,5,6,8}$   & 001 & 000011  \\  
 		$X_h$ & $X_{4,6,8}$     & 001 & 001111  \\    
 		$X_i$ & $X_{4,6}$       & 001 & 001100  \\    
 		$X_j$ & $X_4$           & 001 & 001000  \\
 		\hline
 	\end{tabular}
 	\caption{The syndrome table of the syndrome extraction circuit for the [[9,1,3]] code in Part (B) of Fig.~(\ref{fig:913_parallel}) when 
 		the flag qubit raises.
 		$m_7m_8f_1$ denotes the measurement outcomes in Part (B) and $m'_{123456}$ denotes the measurement outcomes of a following separate unflagged syndrome extraction circuit.
 	}\label{table:error_list_913_p}
 \end{table}

\begin{figure*}[htbp]
	\centering
	\includegraphics[width=1\linewidth]{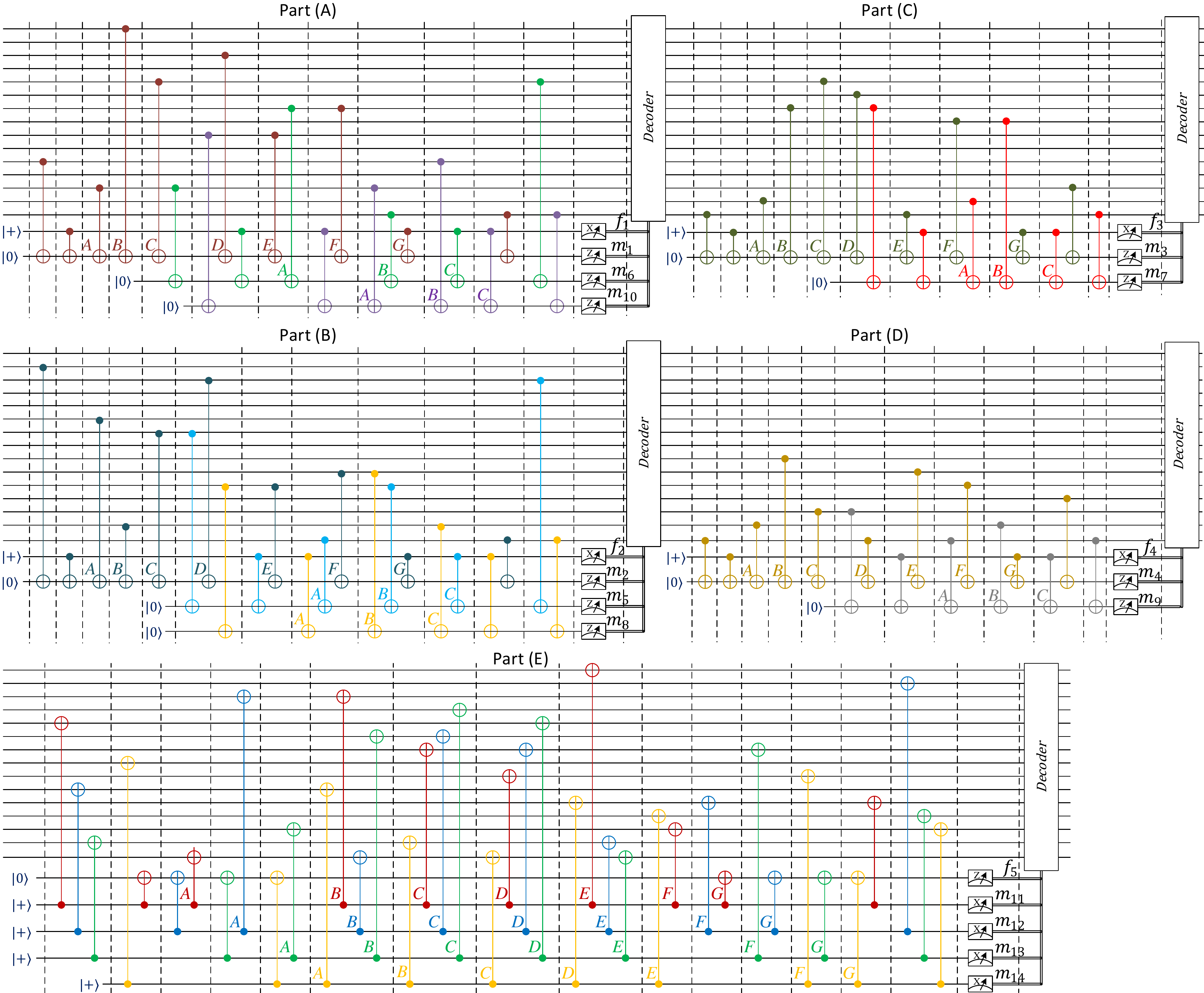}
	\caption{Circuit of the [[15,1,3]] code of parallel extraction scheme. Part (A) simultaneously extracts the syndrome for $g_1$,$g_6$, and $g_{10}$. Part (B) simultaneously extracts the syndrome for $g_2$, $g_5$, and $g_{8}$. Part (C) simultaneously extracts the syndrome for $g_3$, and $g_{7}$. Part (D) simultaneously extracts the syndrome for $g_4$, and $g_{9}$. Part (E) simultaneously extracts the syndrome for $g_{11}$, $g_{12}$, $g_{13}$ and $g_{14}$. Note that the different colors correspond to different stabilizers.}\label{fig:1513_parallel}
\end{figure*}

\subsubsection{The [[15,1,3]] Reed-Muller code}
The [[15,1,3]] Reed-Muller code is defined by stabilizers
\begin{equation*}
\begin{aligned}
g_1   = &Z_{1,3,5,7,9,11,13,15},\, g_6 = Z_{5,7,13,15},\,g_{10}= Z_{9,11,13,15}& \\
g_2   = &Z_{2,3,6,7,10,11,14,15},\, g_5 = Z_{3,7,11,15},\,g_8 = Z_{10,11,14,15} ,& \\
g_3   = &Z_{4,5,6,7,12,13,14,15},\, g_7   = Z_{6,7,14,15},&\\
g_4   = &Z_{8,9,10,11,12,13,14,15},\,g_9= Z_{12,13,14,15},&\\
g_{11}= &X_{1,3,5,7,9,11,13,15},\, g_{12}= X_{2,3,6,7,10,11,14,15} ,&\\
g_{13}= &X_{4,5,6,7,12,13,14,15},\, g_{14}= X_{8,9,10,11,12,13,14,15}.&
\end{aligned}
\end{equation*}

There are ten $Z$-stabilizers $g_1,\dots,g_{10}$ and four $X$-stabilizers $g_{11},\dots, g_{14}$ of weight four or eight.
By Propositions~\ref{pro:count} and~\ref{pro:condition_parallel}, we can have at most three $Z$-stabilizers measured in parallel with one shared flag qubit,
whereas the four $X$-stabilizers can be easily measured in parallel with one shared flag qubit. 
The resulting parallel syndrome extraction circuit is shown in Fig.~\ref{fig:1513_parallel} where Part (A) measures $g_1$, $g_6$ and $g_{10}$, Part (B) measures $g_2$, $g_5$ and $g_8$, Part (C) measures $g_3$ and $g_7$, Part (D) measures $g_4$ and $g_9$, and Part (E) measures $g_{11}$, $g_{12}$, $g_{13}$ and $g_{14}$. Each stabilizer measurement in each part is differently colored. 
One can verify that each bad location failure has a unique error syndrome when a flag rises, and the syndrome table is omitted.

\begin{algorithm}[ht]\label{alg:dec_condition_CSS}
	
		Suppose that a  CSS code of distance three has $X$-stabilizers $g_1^X,\dots,g_x^X$ and $Z$-stabilizers $g_1^Z,\dots,g_z^Z$\;

		$f_X$ is initialized to $0^r$, where $r$ is the number of flag qubits in $C(g_1^X),\dots,C(g_x^X)$\; 
		Apply $C(g_1^X),\dots,C(g_x^X)$ to get the $Z$ error syndrome bits $m_X\in\{0,1\}^x$\; 
		Update $f_X$ according to the measurement outcomes of the flag qubits in $C(g_1^X),\dots,C(g_x^X)$\;
		\uIf{$f_X\neq 0^r$}{
			Apply a separate unflagged syndrome extraction circuit for the $Z$-stabilizers to get syndrome bits $m'_Z\in\{0,1\}^z$\;
				Perform  error correction according to  $\TB(f)$ and $m'_Z$\; 
			
		}
		
		\uElseIf{$f_X=0^r$ and $m_X\neq 0^x$}{
			Apply a complete unflagged syndrome extraction circuit to get syndrome bits $m'\in\{0,1\}^{x+z}$\;
			Perform  error correction according to  $\TB(0)$ and $m'$\; 
		}
		\ElseIf{$f_X=0^r$ and $m_X=0^x$} {
			$f_Z$ is initialized to $0^s$, where $s$ is the number of flag qubits in $C(g_1^Z),\dots,C(g_z^Z)$\; 
			Apply $C(g_1^Z),\dots,C(g_z^Z)$ to get the $X$ error syndrome $m_Z\in\{0,1\}^z$\;
			
			\uIf{$f_Z\neq 0^s$}{
				Apply a separate unflagged syndrome extraction circuit for the $X$-stabilizers to get syndrome bits $m'_X\in\{0,1\}^x$\;
					Perform  error correction according to  $\TB(f)$ and $m'_X$\; 
			}
			\ElseIf{$m_Z\neq 0^z$}{
				Apply a separate unflagged syndrome extraction circuit for the $Z$-stabilizers to get syndrome bits $m'_Z\in\{0,1\}^z$\;
				Perform  error correction according to  $\TB(0)$ and $m'_Z$\; 
			}
		}
	
	\caption{ Decoding procedure for flagged syndrome extraction of a CSS code.}
\end{algorithm}

\begin{algorithm}[htbp]\label{alg:dec_symmetric_CSS}
	
		Suppose a that CSS code of distance three has $X$-stabilizers $g_1^X,\dots,g_x^X$ and $Z$-stabilizers $g_1^Z,\dots,g_z^Z$\;

		$f_X$ is initialized to $0^r$, where $r$ is the number of flag qubits in $C(g_1^X),\dots,C(g_x^X)$\; 
		Apply $C(g_1^X),\dots,C(g_x^X)$ to get the $Z$ error syndrome bits $m_X\in\{0,1\}^x$\; 
		Update $f_X$ according to the measurement outcomes of the flag qubits in $C(g_1^X),\dots,C(g_x^X)$\;
		\uIf{$f_X\neq 0^r$}{
			Apply a separate unflagged syndrome extraction circuit for the $Z$-stabilizers to get syndrome bits $m'_Z\in\{0,1\}^z$\;
			Perform  error correction according to  $\TB(f)$ and $m'_Z$\; 
			
		}
		
		\ElseIf{$f_X\neq 0^r$ and $m_X\neq 0^x$}{
			Apply  a separate unflagged syndrome extraction circuit to get syndrome bits $m_X'\in\{0,1\}^{x}$\;
			Perform  error correction according to  $\TB(0)$ and $m_X'$\; 
		}

		$f_Z$ is initialized to $0^s$, where $s$ is the number of flag qubits in $C(g_1^Z),\dots,C(g_z^Z)$\; 
		Apply $C(g_1^Z),\dots,C(g_z^Z)$ to get the $X$ error syndrome $m_Z\in\{0,1\}^z$\;
		
		\uIf{$f_Z\neq 0^s$}{
			Apply a separate unflagged syndrome extraction circuit for the $X$-stabilizers to get syndrome bits $m'_X\in\{0,1\}^x$\;
			Perform  error correction according to  $\TB(f)$ and $m'_X$\; 
		}
		
		\ElseIf{$f_X\neq 0^r$ and $m_Z\neq 0^z$}{
			Apply a separate unflagged syndrome extraction circuit for the $Z$-stabilizers to get syndrome bits $m'_Z\in\{0,1\}^z$\;
			Perform  error correction according to  $\TB(0)$ and $m'_Z$\;
		}

	\caption{(Symmetric) Decoding procedure for flagged syndrome extraction of a CSS code.}
\end{algorithm}

\subsection{Decoding procedures}
Finally, we propose a decoding procedure for flagged syndrome extraction of a CSS code.  We begin with the following lemma.
\bl \label{lemma:CSS_error}
A location failure in a flagged syndrome extraction circuit of an $X$-stabilizer (respectively, $Z$-stabilizer)  will leave solely $X$  (respectively, $Z$) errors 
and  possibly a $Z$ (respectively, $X$) error on the data qubits.

\el
\noindent\textbf{Proof. }
Consider the flagged syndrome extraction circuit in Fig.~\ref{fig:flagged_syndrome_extraction}.
Consider a location failure at one of the CNOTs that triggers the flag qubit.
It may generate a Pauli error on the control (data) qubit and a $Z$ error on the target (ancillary) qubit, and this $Z$ error will propagate to multiple data qubits. 
The condition is similar for the syndrome extraction of a $Z$-stabilizer. \\

Note that in Fig.~\ref{fig:flagged_syndrome_extraction},  a bad location failure will trigger the flag qubit, and the high-wight $Z$ error can be detected by measuring the $X$-stabilizers. Meanwhile, a possible residual $X$ error can be detected and corrected by measuring the $Z$-stabilizers.

We remark that this residual $X$ error can be left uncorrected  to the next error correction cycle, whereas the syndrome extraction procedure remains to be fault-tolerant.  This suggests a decoding procedure for CSS codes as shown in~Algorithm~\ref{alg:dec_condition_CSS}, 
where  separate unflagged syndrome extractions for $X$ and $Z$ stabilizers are applied, respectively.
Since we do not perform a complete syndrome extraction, the overhead for FTQC can be greatly reduced.

We remark that $X$ and $Z$ errors are differently treated in  Algorithm~\ref{alg:dec_condition_CSS}. In a sequence of error correction cycles, we may alternate the order of $X$ and $Z$ treatments in each cycle.

As a comparison, we also provide Algorithm~\ref{alg:dec_symmetric_CSS}, where $X$ and $Z$ errors are symmetrically treated.
These two decoding algorithms will be compared in the next section, where we show by simulations that Algorithm~3 performs slightly better.

\section{Simulation results}\label{sec:sim}

In this section, we simulate the memory and computation thresholds of the error detection scheme based on the [[4,2,2]] code with standard and parallel flagged syndrome extraction, respectively,
the scheme based on the [[7,1,3]] Steane code with parallel syndrome extraction, and the scheme based on Shor's [[9,1,3]] code with parallel syndrome extraction.

We consider depolarizing errors.  We assume that  all the gates have the same error rate and 
the ratio of memory error rate  (idle qubits) to the gate error rate is $\gamma$. Simulations will be conducted at $\gamma=0$ and $\gamma=1$.

The number of locations in an ex-Rec CNOT gate of each scheme is provided in Table \ref{table:error_number}. 
Note that the $[[4,2,2]], [[7,1,3]],$ and $[[9,1,3]]$ flag schemes follow the standard flagged syndrome extraction in Algorithm~\ref{algorithm:standard};
the [[4,2,2]], [[7,1,3]] parallel schemes are given in Ref.~\cite{Rei20};
the [[9,1,3]] parallel scheme is proposed  in Section~\ref{sec:913} with   { the decoding procedure in Algorithm~\ref{alg:dec_condition_CSS}.} Our code can be found in Ref.~\cite{parallel_sf}.

\begin{table}[htbp]
	\begin{tabular}{|c|c|c|c|c|c|c|} 
			\hline
			& [[4,2,2]]  & [[4,2,2]]    &[[7,1,3]]  & [[7,1,3]]       &  [[9,1,3]]  & [[9,1,3]]  \\
			& Flag       & \footnotesize Parallel     & Flag $^\text{a}$  &  \footnotesize Parallel $^\text{a}$   &  Flag $^\text{a}$   & \footnotesize Parallel \footnote{These numbers are overestimated since the locations in all the unflagged syndrome extraction circuits are counted.}       \\
			\hline
			${P_{\text{s}}}$ \footnote{$P_s$ denotes ancilla state preparation.}        & 16  		& 	8	   		& 	264  	&  72    & 176 & 100 \\
			\hline 
			$M_X$                      & 8   		& 	4	   		& 	132   	&  36    & 24    & 24\\
			\hline
			$M_Z$                      & 8   		& 	4   		& 	132   	&  36    & 152   & 76\\
			\hline
			CNOT                       & 52  		& 	36  		& 	1015  	&  343   & 457  & 313\\
			\hline
			\footnotesize Idle qubit   & 192 		& 	64  		& 	6360 	&  824   & 1688  & 1284 \\
			\hline
			SWAP                       & 0  		& 	8  			& 	0    	&   0    & 0    & 0\\
			\hline
			Total                      & 276 		& 	124 		& 	7903 	&  1311  & 2497 & 1797\\
			\hline
		\end{tabular}
	
	\caption{The number of locations in an ex-Rec CNOT gate for various schemes. 
		}\label{table:error_number}

\end{table}

\begin{figure*}[htbp]
	\centering
	\includegraphics[width=1\linewidth]{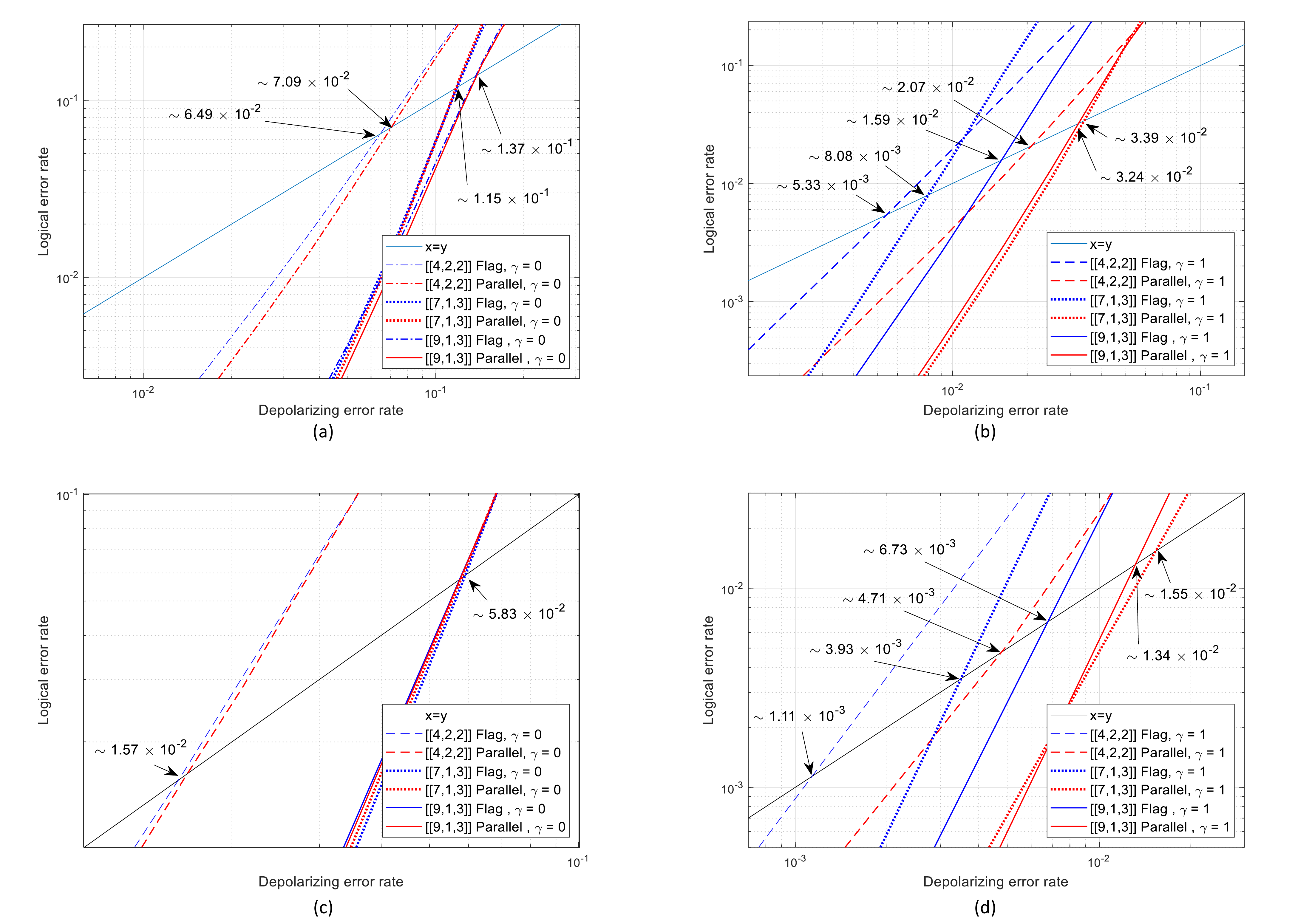}
	 	\caption{Error correction: simulations of the memory and computation pseudo-thresholds for various  schemes  at $\gamma=0$ and  $\gamma=1$, using Algorithm~\ref{alg:dec_condition_CSS}. (a) Memory pseudo-thresholds at $\gamma=0$. (b) Memory pseudo-thresholds at $\gamma=1$. (c) Computation pseudo-thresholds at $\gamma=0$. (d) Computation pseudo-thresholds at $\gamma=1$. }\label{simulations_det}
\end{figure*}

 \subsection{Comparisons of procedures for unflagged syndrome extraction}
We proposed Algorithms~\ref{alg:dec_condition_CSS} and~\ref{alg:dec_symmetric_CSS} for CSS codes so that a complete unflagged syndrome extraction is not required for CSS codes when some flag qubits rise.
 In this subsection, we demonstrate this on the parallel flagged syndrome extraction for the error correction scheme based on the [[9,1,3]] code proposed  in Section~\ref{sec:913}.

The memory and computation pseudo-thresholds using  the decoding procedure in Algorithm~\ref{alg:dec_symmetric_CSS}  with complete unflagged syndrome extraction or  
 Algorithms~\ref{alg:dec_condition_CSS}  and~\ref{alg:dec_symmetric_CSS} with separate unflagged syndrome extraction are summarized in Table~\ref{tb:CSS_comp}.
The results show that using a separate and nonsymmetric unflagged syndrome extraction by Algorithm~\ref{alg:dec_condition_CSS} is better   in all respects. Consequently, we will adopt this strategy in the following simulations.

 \begin{table}[ht]
 	
 	\begin{center}
 		\begin{tabular}[t]{|l|c|c|}

 			\hline
 			  $[[9,1,3]]$ Parallel Scheme      &      $\gamma = 0$                & $\gamma = 1$        \\
 			\hline               
 			  Decoding Procedure           		& \multicolumn{2}{c|}{Memory pseudo-threshold }      \\
 			\hline
  			 Alg.~\ref{alg:dec_symmetric_CSS} with comp.		&  $ 8.01 \times 10^{-3}$      & $  8.3 \times 10^{-4}$ \\
 		 Alg.~\ref{alg:dec_symmetric_CSS} 	&  $ 8.06 \times 10^{-3}$      & $8.52 \times 10^{-4}$  \\
 		 Alg.~\ref{alg:dec_condition_CSS}		&  $9.82 \times 10^{-3}$      & $8.84 \times 10^{-4}$ \\
 			\hline \hline
 	           Decoding Procedure       &  \multicolumn{2}{c|}{ Computation pseudo-threshold}  \\
 			\hline
  		 Alg.~\ref{alg:dec_symmetric_CSS} with comp.	&  $ 3.94 \times 10^{-4}$      & $  3.35 \times 10^{-5}$  \\
  		 Alg.~\ref{alg:dec_symmetric_CSS} 	&  $4.19 \times 10^{-4}$      & $4.97 \times 10^{-5}$  \\
 			Alg.~\ref{alg:dec_condition_CSS}		&  $7.81 \times 10^{-4}$      & $8.11 \times 10^{-5}$ \\
 			\hline
 		\end{tabular}
 		\caption{Comparisons of the three procedures for unflagged syndrome extraction. Note that Alg.~\ref{alg:dec_condition_CSS}  and~Alg.~\ref{alg:dec_symmetric_CSS} are the  $X$ and $Z$ separate unflagged syndrome extraction, while Alg.~\ref{alg:dec_symmetric_CSS} with comp. uses a complete unflagged syndrome extraction instead.  }\label{tb:CSS_comp}
 	\end{center}
  	
 \end{table}

\subsection{Error detection}
Simulations of the [[4,2,2]], [[7,1,3]], and [[9,1,3]] codes with error detection are given in Fig.~\ref{simulations_det}.
 The $[[4,2,2]], [[7,1,3]],$ and $[[9,1,3]]$ flag schemes are with standard flagged syndrome extraction;
the [[4,2,2]], [[7,1,3]] parallels schemes are given in \cite{Rei20} and can be found in Appendix~\ref{sec:parallel_flag};
the [[9,1,3]] Parallel is the scheme proposed  in Section~\ref{sec:913}.

Figures~\ref{simulations_det} (a) and (b) provide the memory pseudo-thresholds at $\gamma=0$ and $\gamma=1$, respectively,
while Figures~\ref{simulations_det} (c) and (d) provide the computation pseudo-thresholds at $\gamma=0$ and $\gamma=1$, respectively.

In general, the flag and parallel schemes have comparable pseudo-thresholds at $\gamma=0$.
At $\gamma=1$, the advantages of parallel scheme is more obvious as expected since  it has fewer qubits and gates involved.

At $\gamma=0$, both the [[7,1,3]] and [[9,1,3]] codes have computation pseudo-thresholds around $5\%$,
while their memory pseudo-thresholds are above $10\%$.
At $\gamma=1$, the [[7,1,3]] parallel is slightly better than the [[9,1,3]] parallel scheme for error detection.

\subsection{Error correction}

\begin{figure*}[htbp]
	\centering
	\includegraphics[width=1\linewidth]{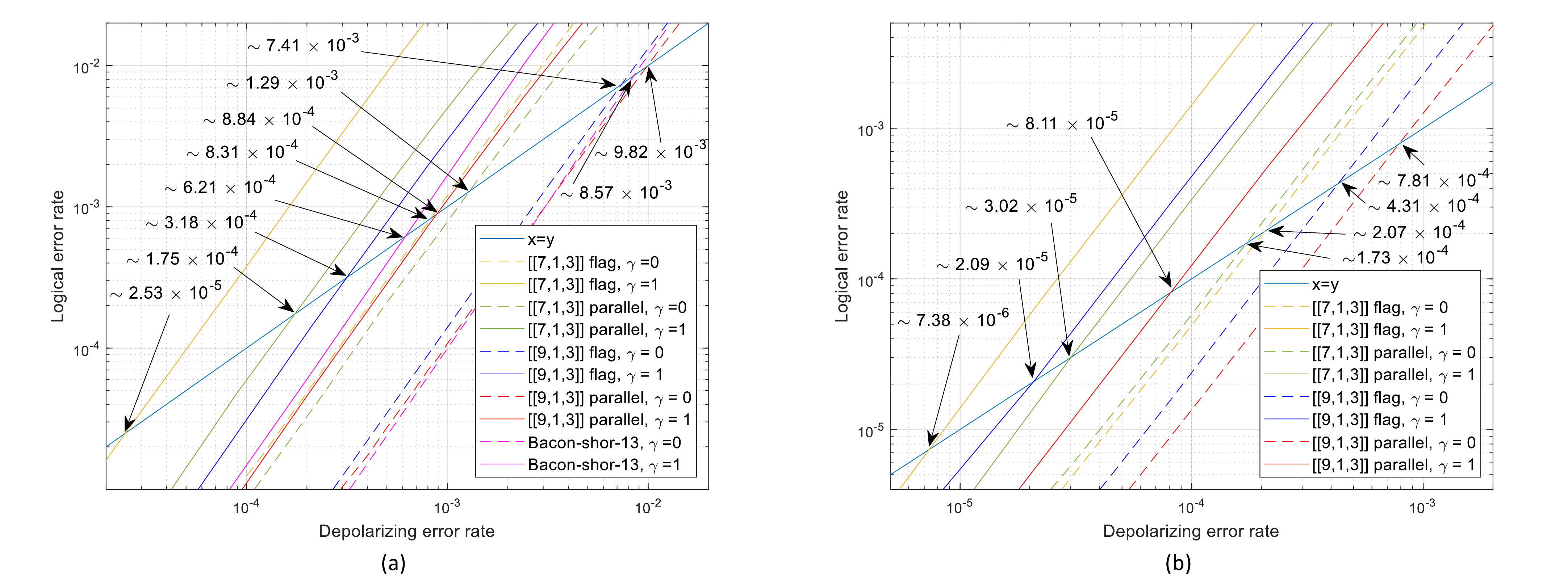}
	 	\caption{Error correction: simulations of the memory and computation pseudo-thresholds for various  schemes  at $\gamma=0$ and  $\gamma=1$, using Algorithm~\ref{alg:dec_condition_CSS}. (a) Memory pseudo-thresholds. (b) Computation pseudo-thresholds. }\label{simulations_ecc}
\end{figure*}

Simulations of the  [[7,1,3]] and [[9,1,3]] codes with error correction are given in Fig.~\ref{simulations_ecc}.

We first consider the memory  pseudo-threshold of the [[7,1,3]] and [[9,1,3]] schemes using  {Algorithm~\ref{alg:dec_condition_CSS}} for error correction, which are shown in Fig.~\ref{simulations_ecc}~(a) and (b), respectively.
The memory and computation pseudo-thresholds for these schemes using  {Algorithm~\ref{alg:dec_condition_CSS}} are summarized in Table~\ref{tb:std_result}.

As a comparison, we include the pseudo-threshold of the [[7,1,3]] flag scheme using a two-round decoder in~\cite{CB18}, which has a better threshold than the [[7,1,3]] flag scheme using the standard decoder. 

\begin{figure}[htbp]
	\centering
	\includegraphics[width=1\linewidth]{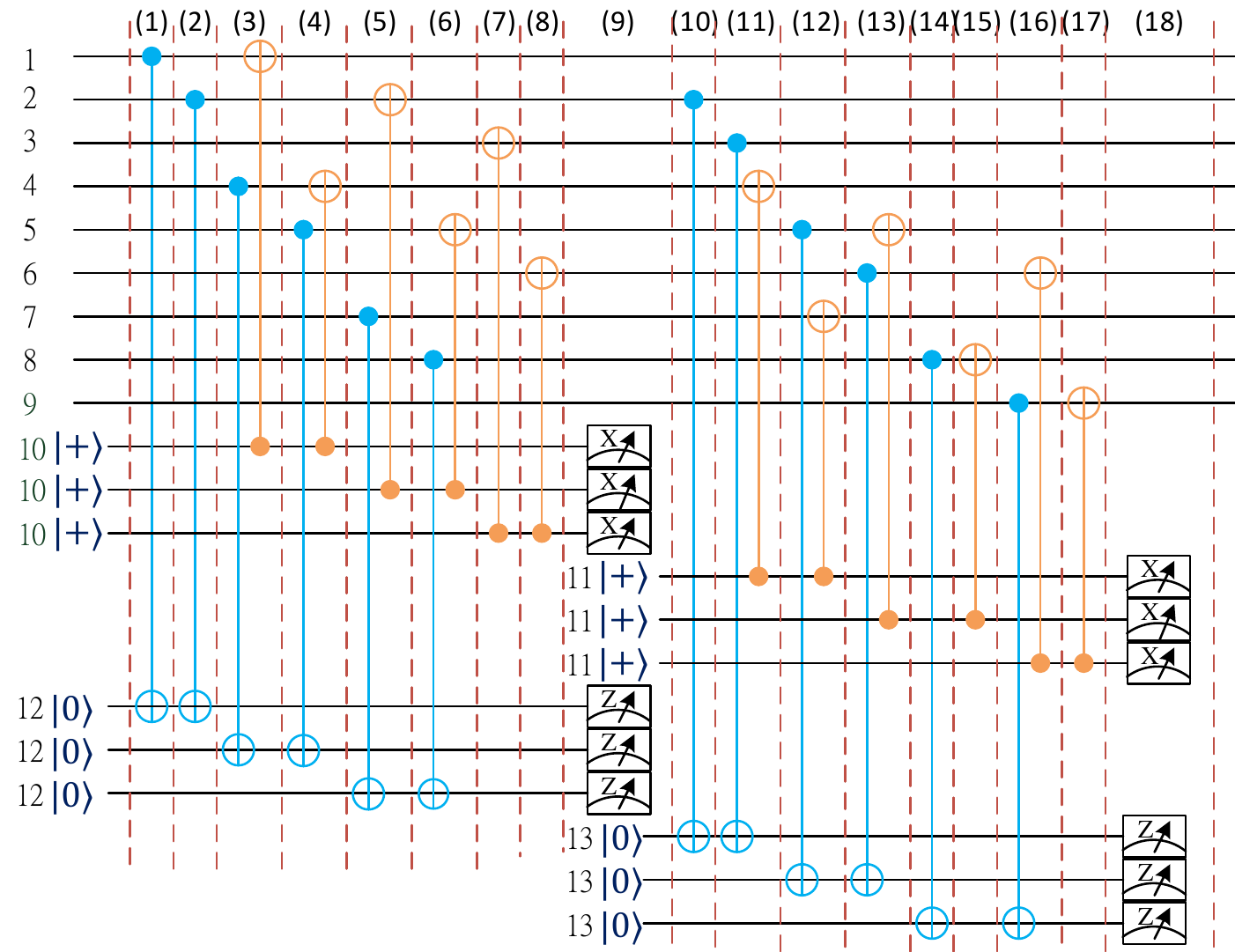}
	\caption{The circuit for the Bacon-Shor-13 scheme.}\label{fig:BS13}
\end{figure}

We also compare our [[9,1,3]] parallel scheme with the \textit{Bacon-Shor-13} scheme in Ref.~\cite{LMB18}.
For a fair comparison, we reproduce their results and simulate the memory pseudo-threshold for the Bacon-Shor-13 scheme at $\gamma=0$ and $\gamma=1$
based on the circuit for Bacon-Shor-13 in Fig.~\ref{fig:BS13} and the results are also shown in Table~\ref{tb:std_result} and Fig.~\ref{simulations_ecc}\,{(a)}.

It is obvious from the simulations  that the parallel schemes are much better than their flag counterparts
since the parallel schemes have fewer qubits and gates as can be observed from  Table~\ref{table:error_number}.

Remarkably, our [[9,1,3]] parallel scheme has the highest  memory pseudo-thresholds of
 $9.82 \times 10^{-3}$ and $8.84 \times 10^{-4}$ at $\gamma=0$ and $\gamma=1$, respectively, among these schemes.

The simulation results for the computation pseudo-thresholds using  {Algorithm~\ref{alg:dec_condition_CSS}} have similar behaviors as shown in Fig.~\ref{simulations_ecc}~(b).
Our [[9,1,3]] parallel scheme also outperforms the [[7,1,3]] parallel scheme in the computation pseudo-threshold.

\begin{table}[htbp]

	\begin{center}
		\begin{tabular}[t]{|l|c|c|}
			\hline
			  & $\gamma = 0$                & $\gamma = 1$           \\
			\hline               
			Scheme               &     \multicolumn{2}{c|}{Memory pseudo-threshold }    \\
			\hline
			$[[7,1,3]]$ flag     &  $8.31 \times 10^{-4}$      & $2.53 \times 10^{-5}$  \\
			$[[7,1,3]]$ flag      &            ---             & $3.39 \times 10^{-5}$ \cite{CB18} \footnote{A two-round decoder is used.}   \\
		    $[[7,1,3]]$ parallel \cite{Rei20} &  $1.29 \times 10^{-3}$      & $1.75 \times 10^{-4}$ \\
			$[[9,1,3]]$ flag     &  $7.41 \times 10^{-3}$      & $3.18 \times 10^{-4}$  \\
			$[[9,1,3]]$ parallel &  $9.82 \times 10^{-3}$      & $8.84 \times 10^{-4}$  \\
			Bacon-Shor-13 \cite{LMB18} &    $8.57 \times 10^{-3}$  \cite{LMB18}  & $6.21 \times 10^{-4}$     \cite{LMB18}     \\  
			\hline \hline
			Scheme               &  \multicolumn{2}{c|}{ Computation pseudo-threshold}  \\
			\hline
			$[[7,1,3]]$ flag     & $2.07 \times 10^{-4}$ & $7.38 \times 10^{-6}$  \\
			$[[7,1,3]]$ parallel \cite{Rei20} & $1.73 \times 10^{-4}$ & $3.02 \times 10^{-5}$  \\
			$[[9,1,3]]$ flag     & $4.31 \times 10^{-4}$ & $2.09 \times 10^{-5}$  \\
			$[[9,1,3]]$ parallel & $7.81 \times 10^{-4}$ & $8.11 \times 10^{-5}$  \\   
			\hline
		\end{tabular}
		\caption{Memory and computation pseudo-thresholds for various schemes of the $[[7,1,3]]$ and $[[9,1,3]]$ codes.}\label{tb:std_result}
	\end{center}

\end{table}

\section{Conclusion}\label{sec:conclusion}
We proposed a general parallel flagged syndrome extraction scheme for a CSS code of distance three with one shared flag qubit (Algorithm~\ref{algorithm:shared_flag}),
together with a decoding procedure for  separate  unflagged syndrome extraction (Algorithm~\ref{alg:dec_condition_CSS}).
Examples of the [[9,1,3]] Shor and [[15,1,3]] Reed-Muller codes were provided. 

Simulations on the  [[9,1,3]] code demonstrated that our [[9,1,3]] parallel scheme has several benefits over the [[7,1,3]] code.
Also it has a high memory pseudo-threshold around $8.84 \times 10^{-4}$, assuming all the gates and memory have the same error rate.
The [[9,1,3]] parallel scheme uses six ancillas for the $Z$-stabilizer measurements, where three ancillas can be reset for the $X$-stabilizer measurements in the second part. 
It is also better than Bacon-Shor-13 at the cost of two additional qubits for parallelism.
Since Bacon-Shor-13 uses only four ancilla qubits for syndrome extraction, it has long circuit depth and many idle qubit locations.

Our procedure works for CSS codes. However, there are examples of non-CSS codes with shared flag qubits in Refs.~\cite{CR18a,Rei20}.
A procedure for a general stabilizer code maybe be possible but it is beyond our scope.

	In Ref.~\cite{LA20}, Lao and Almudever~proposed a two-dimensional layout for the [[7,1,3]] Steane code with a shared flag qubit~\cite{CR18a}.
	In particular, they consider a variant of the standard flagged syndrome extraction to reduce the circuit depth.
Consequently, the CNOT gates for a stabilizer measurement may be equally connect to an ancilla qubit and the flag qubit.
The circuit depth can be decreased in this way. However, it may not be easy to have parallel syndrome extraction for multiple stabilizers with one shared flag qubit since it will be difficult to find an arrangement of the CNOT gates such that each bad location failure will have a unique error syndrome. 

 We consider only quantum codes of distance two and three in this paper.
It is possible to extend our framework for CSS codes of higher distances, where more shared flag qubits are required. 
However, it is difficult to arrange the CNOTs so that each bad location failure will have a unique error syndrome by a complete unflagged syndrome extraction.

\section*{Acknowledgments} 
CYL was supported by the National Science and Technology Council (NSTC) in
Taiwan, under Grants No.\,NSTC110-2628-E-A49-007, No. NSTC111-2119-M-A49-004, and No. NSTC111-2119-M-001-002.

\appendix

\section{Parallel flagged syndrome extraction of  the $[[4,2,2]]$ and $[[7,1,3]]$ codes}\label{sec:parallel_flag}

\begin{figure}[htbp]
	\centering
	\includegraphics[width=0.84\columnwidth]{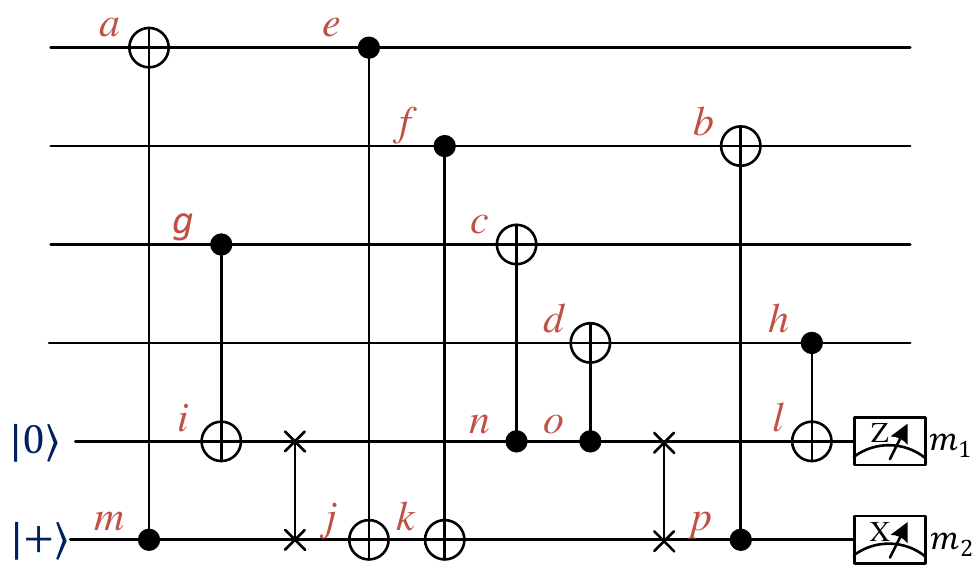}
	\caption{Parallel syndrome extraction\footnote{This circuit is a modification of the circuit in \cite{Rei20} so that the code space is preserved.
		} of $X_1X_2X_3X_4$ and $Z_1Z_2Z_3Z_4$ for the [[4,2,2]] code.}
	\label{fig:Parallel_syndrome}
\end{figure}

We explain Reichardt's parallel syndrome extraction with flag qubits~\cite{Rei20} in the following. 
Figure~\ref{fig:Parallel_syndrome} shows a parallel syndrome extraction circuit for the $[[4,2,2]]$ code, whose circuit depth is lower than the cascade form in Fig.~\ref{fig:Logic CNOT_flag}.
Only two ancilla qubits are used  to extract two syndrome bits, and each of them serves as a flag qubit for the other. Note that two SWAP gates are introduced, instead of CNOT gates.

Each single location failure in the parallel circuit  Fig.~\ref{fig:Parallel_syndrome}  and its resulting data error are provided in Table~\ref{tb:errorlist_422}, where $X_a$ or $Z_a$ denotes a corresponding Pauli error at location $a$.
A few notes on the table are as follows.
The ancilla failures, such as $X_i$ and $Z_m$ do not introduce any data qubit errors; however, they cannot be identified from the syndromes and the uncorrupted codewords have to be discarded. On the other hand, $X_m$ and $Z_i$ induce certain data errors that are stabilizer generators and hence the data qubits are not affected. Certain failures, such as $X_p$ and $Z_l$ induce weight-1 errors with zero syndrome, but these errors will be detected and corrected in the next round of error detection.

\begin{table}
	\begin{tabular}{|c|c|c|c|c|c|}
		\hline
		Failure & Data error & $m_{1,2}$ &  Failure & Data error & $m_{1,2}$ \\
		\hline
		$X_a$ & $X_1$         & 10 & $Z_a$ & $Z_1$         & 01 \\
		$X_b$ & $X_2$         & 00 & $Z_b$ & $Z_2$         & 01 \\
		$X_c$ & $X_3$         & 00 & $Z_c$ & $Z_3$         & 01 \\
		$X_d$ & $X_4$         & 10 & $Z_d$ & $Z_4$         & 01 \\
		$X_e$ & $X_1$         & 10 & $Z_e$ & $Z_1$         & 00 \\
		$X_f$ & $X_2$         & 10 & $Z_f$ & $Z_2$         & 01 \\
		$X_g$ & $X_3$         & 10 & $Z_g$ & $Z_3$         & 01 \\
		$X_h$ & $X_4$         & 10 & $Z_h$ & $Z_4$         & 00 \\
		$X_i$ & None      & 10 & $Z_i$ & $Z_{1,2,3,4}$ & 00 \\
		$X_j$ & None       & 10 & $Z_j$ & $Z_{1,2,4}$   & 01 \\
		$X_k$ & None       & 10 & $Z_k$ & $Z_{2,4}$     & 01 \\  
		$X_l$ & None       & 10 & $Z_l$ & $Z_4$         & 00 \\  
		$X_m$ & $X_{1,2,3,4}$ & 00 & $Z_m$ & None      & 01 \\
		$X_n$ & $X_{2,3,4}$   & 10 & $Z_n$ & None      & 01 \\
		$X_o$ & $X_{2,4}$     & 10 & $Z_o$ & None      & 01 \\
		$X_p$ & $X_2$         & 00 & $Z_p$ & None      & 01 \\
		\hline
	\end{tabular}
	\caption{Equivalent errors on the data qubits for various single location failures and their syndromes in  the parallel syndrome extraction circuit of $X_1X_2X_3X_4$ and $Z_1Z_2Z_3Z_4$ in Fig.~\ref{fig:Parallel_syndrome} for the [[4,2,2]] code.}\label{tb:errorlist_422}
\end{table}

Next we explain parallel syndrome extraction with shared flag qubits for the [[7,1,3]] Steane code~\cite{Ste96a},
which is  defined by the following stabilizer generators
\begin{align}
	\begin{array}{lll}
		X_1X_3X_5X_7,& Z_2Z_3Z_6Z_7,&Z_4Z_5Z_6Z_7,\\
		Z_1Z_3Z_5Z_7,&X_2X_3X_6X_7,& X_4X_5X_6X_7.
	\end{array} \label{eq:Steane_stabilizers}
\end{align}

 \begin{figure}[hptb]
 	\centering
 	\includegraphics[width=1\columnwidth]{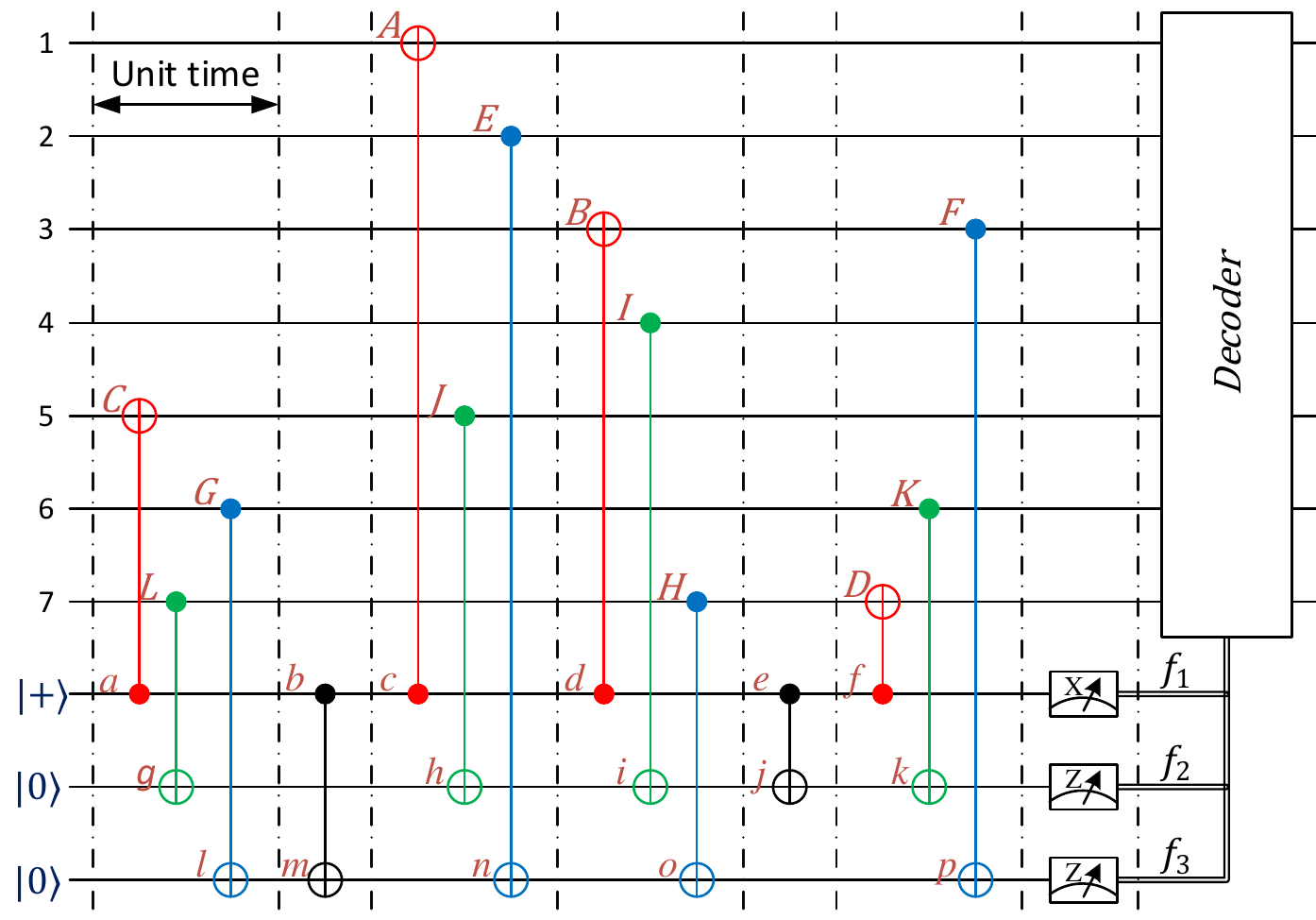}
 	\caption{Parallel syndrome extraction circuits for the $[[7,1,3]]$ code. (A):  $X_1X_3X_5X_7$ (red), $Z_2Z_3Z_6Z_7$ (blue), and $Z_4Z_5Z_6Z_7$ (green).  (B):  $Z_1Z_3Z_5Z_7$, $X_2X_3X_6X_7$, and $X_4X_5X_6X_7$. }\label{fig:713_parallel}
 \end{figure}

Reichardt demonstrated   that two or three stabilizer generators in Eq.~(\ref{eq:Steane_stabilizers}) can be simultaneously measured in a fault-tolerant way~\cite{Rei20}.
Figure~(\ref{fig:713_parallel}) illustrates the case of measuring three stabilizers  $X_1X_3X_5X_7$, $Z_2Z_3Z_6Z_7$, and $Z_4Z_5Z_6Z_7$  in parallel and only three ancillas are required~\cite{Rei20}.
The two additional CNOTs connecting the ancillas are elegantly placed so that each of the ancillas can serve as a flag qubit for the others.

 Unlike our protocol, a complete unflagged syndrome extraction circuit is applied here  if anyone of the binary measurement outcomes $f_1,f_2,f_3$ is~$1$. 
To reduce the circuit depth, one can apply an unflagged version of Fig.~(\ref{fig:713_parallel})
to extract the three syndrome bits and similarly for the other three.

	The idea of shared flag qubits also works for the $[[5,1,3]]$ code and the $[[15,7,3]]$ CSS code so that
	two stabilizers can be simultaneously measured but an extra flag qubit is necessary~\cite{Rei20}.

\bibliography{qecc}

\end{document}